\documentclass[a4paper]{jpconf}
\usepackage{graphicx}
\usepackage{amsfonts}
\usepackage{amsmath}
\usepackage{amssymb}
\usepackage{dsfont}

\renewcommand{\Tr}{\mathop{\mathrm{Tr}}\nolimits}

\begin{document}
\hfill {\bf HU-EP-15/45} \\

\vspace*{-2cm}
\title{Topological susceptibility from $N_f=2+1+1$ lattice QCD at nonzero temperature
\footnote[99]{Contribution to the International Conference 
{\it Strangeness in Quark Matter 2015}, Dubna, July 6 -- 11, 2015, presented by
A. Trunin}}

\author{Anton~Trunin$^{1}$, Florian~Burger$^{2}$, Ernst--Michael~Ilgenfritz$^{1}$, \\
Maria~Paola~Lombardo$^{3}$ and Michael~M\"uller--Preussker$^2$}

\address{$^1$ Bogoliubov Laboratory of Theoretical Physics, Joint Institute for Nuclear Research, \\
$~$ Joliot--Curie Str. 6, 141980 Dubna, Russia}

\address{$^2$ Physics Department, Humboldt University Berlin, Newtonstr. 15, 12489 Berlin, Germany}

\address{$^3$ Frascati National Laboratory, National Institute for Nuclear Physics, \\ 
$~$ Via Enrico Fermi 40, 00044 Frascati (Rome), Italy}

\ead{amtrnn@gmail.com}

\begin{abstract}
We present results for the topological susceptibility at nonzero temperature 
obtained from lattice QCD with four dynamical quark flavours. We apply different 
smoothing methods, including gradient Wilson flow and over--improved cooling, 
before calculating the susceptibility. 
It is shown that the considered smoothing techniques basically agree 
among each other, and that there are simple scaling relations between flow 
time and the number of cooling/smearing steps. The topological susceptibility 
exhibits a surprisingly slow decrease at high temperature.
\end{abstract}

The non--trivial topological structure of gauge fields and the computation of the
topological susceptibility $\chi_{top}$ is discussed in lattice QCD
since long time. Recent considerations (see~\cite{mueller-preussker14} and 
references therein) ranging from the restoration of the
$U_A(1)$ symmetry at high temperature (or density) to the abundance of 
cosmic axions~\cite{berkowitz} are calling for a better knowledge of 
$\chi_{top}$ as a function of the temperature and quark masses.  

In this study we calculate $\chi_{top}$ in the temperature range 
$150 < T <500$~MeV using lattice configurations generated 
with $N_f=2+1+1$ dynamical Wilson twisted mass fermions at finite temperature 
~\cite{burger13}.  The heavy doublet of $s$ and $c$ quarks has its 
mass parameters matching the physical $K$ and $D$ meson masses, 
while the HMC simulations still require light quarks being unnaturally heavy. 
The configurations are generated at three coupling values $\beta=1.90$, 1.95, and 2.10
(for later continuum extrapolations) with the Wilson twisted mass fermionic 
action tuned to maximal twist, taking benefit from automatic $\mathcal O(a)$ 
improvement. Several charged pion masses are considered, but here we restrict 
ourselves to one value $m_{\pi^\pm} \simeq 370$~MeV.

We focus on the comparison between different smoothing 
methods for lattice gauge fields necessary to get $\chi_{top}$ by the gluonic method.
The smoothing techniques under consideration are the Wilson flow~\cite{luescher2}, 
Wilson and over--improved cooling~\cite{perez}, and stout--link smearing~\cite{moran}.
We use the Wilson flow to set two stopping scales determined by
%\vspace{-.3cm}
\begin{equation}
\begin{gathered}
\label{stop}
t^2\langle E\rangle\bigl|_{t=t_0}=0.3 \quad \text{or} 
\quad t^2\langle E\rangle\bigl|_{t=t_1}=0.66,
\qquad
E =\frac1{2N_\tau N_\sigma^3}\sum_x\Tr[F_{\mu\nu}F^{\mu\nu}(x)],
\vspace{-.2cm}
\end{gathered}
\end{equation}
where $F_{\mu\nu}(x)$ is the field strength tensor on the lattice.
We match ensemble averages $\langle E \rangle$ obtained in other 
methods to the values measured with the Wilson flow at $t_0$ and $t_1$
in order to relate these to corresponding numbers of cooling steps.
Empirically we find in a good agreement for both the stopping criteria 
$t_0$ and $t_1$  that  $N_\text{cool}^\text{Wilson} \simeq 3 \tau$~\cite{bonati}, 
$N_\text{cool}^\text{ov.-imp.} \simeq 5 \tau$ 
and $N_\text{smear}^\text{stout} \simeq 12 \tau$, the latter for our choice 
$\rho_\text{smear}=0.06$ ($\tau$ defined by $t=a^2 \tau$, $a$ is the lattice spacing). 

If the diffusion radius $R = \sqrt{8t}$ at $t_0$ or $t_1$ satisfies the condition 
$ R \ll 1/T$ the topological susceptibility $\chi_{top}(T)$ 
can be safely determined as
\vspace{-.18cm}
\begin{equation}
\label{chi-def}
\chi_{top} = \frac{\langle Q_{top}^2 \rangle}{V}, \qquad 
Q_{top}=\frac{a^4}{32\pi^2}e^{\mu\nu\rho\sigma}\sum_x\Tr[F_{\mu\nu}F_{\rho\sigma}(x)],
\qquad V=a^4 N_\tau N_\sigma^3.
\vspace{-.2cm}
\end{equation}

\begin{figure}[t]
\includegraphics{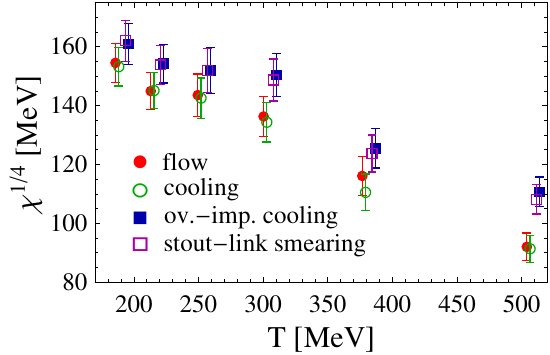}
\hspace{1pt}
\includegraphics{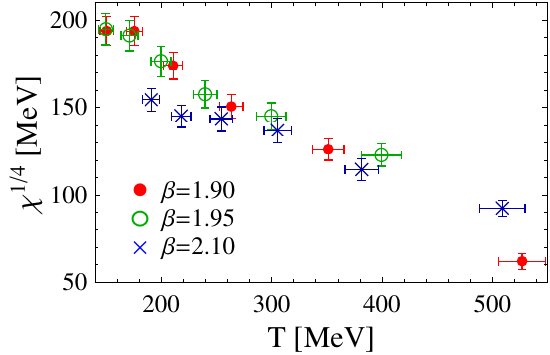}
\hspace{1pt}
\begin{minipage}[b]{10pc}
%\vspace*{-5pt}
\caption{\label{fig} $T$ dependence of $\chi_{top}^{1/4}$ from Wilson flow. 
Left: comparison with all other smoothing methods for $\beta=2.10$ and $t_0$.
Right: at stopping time $t_0$ for three different $\beta$ values. 
}
\end{minipage}
\vspace{-.6cm}
\end{figure}

The spatial and temporal lattice sizes are 
$N_\sigma=24$, 32 and $N_\tau=4 \ldots 16$, respectively, and the statistics 
is varying in the range $200 \ldots 1000$ configurations, depending on the 
respective temperature. One can see from the left panel of Figure~\ref{fig}, 
where the methods are compared for the finest lattice spacing 
(at $\beta=2.10$), that the pairs of Wilson flow and Wilson cooling, 
as well as over--improved cooling and stout--link smearing, give almost 
indistinguishable $\chi_{top}(T)$ values throughout the considered 
temperature range. In the same manner, the agreement between the two 
stopping criteria $t_0$ and $t_1$ can be checked independently for each 
algorithm. The results for $\chi_{top}$ from Wilson and over--improved cooling 
agree (within errors) for low temperatures. Starting from 
approximately $T \gtrsim 300$~MeV, the over--improved result for
$\chi_{top}$ turns out somewhat larger than the results for Wilson flow 
and Wilson cooling as one would expect.

In the right panel of Figure~\ref{fig}, $\chi_{top}^{1/4}(T)$
is shown as calculated at three different $\beta$ values 
(different lattice spacings $a$). The curves beyond $\sim\!200$~MeV can be 
reasonably approximated with linear fits, with slopes visibly flattening 
with the growth of $\beta$ (towards $a \to 0$). Note that the value of 
crossover temperature obtained from the chiral susceptibility is 
$T_c=184(4)$~MeV. Thus, we come to the preliminary conclusion that for 
$N_f=2+1+1$ and $m_{\pi^\pm} \simeq 370$~MeV  the topological susceptibility 
decreases very slowly beyond $T_c$ in the continuum limit, in contrast to the 
rapid fall--off observed in the quenched approximation ($N_f=0$), and 
to the gradual descent taking place in the $N_f=2$ case~\cite{bornyakov}.

We are grateful to the HLRN supercomputer centers Berlin and Hannover, 
the Supercomputing Center of Lomonosov Moscow State University,
and to the HybriLIT group of JINR for computational resources. 
The work was supported by the Heisenberg--Landau program of BLTP JINR 
and BMBF of Germany, and by the ''Dynasty`` foundation.

\end{document}